\documentclass[12pt,aps,pra,showpacs,amsmath,amssymb,tightenlines,floats,floatfix]{revtex4}
\usepackage{mathptm}
\usepackage{hyperref}
\newcommand{\hnl}{\htmladdnormallink}

\begin{document}

\title{The $^{\bf 3}$He$^{\bf 4}$He$_{\bf 2}$
trimer within the hard-core Faddeev approach
\footnote{arXiv: physics/0304048}\footnote{The work
was supported by the Academia Sinica, the National Science
Council (ROC), the Deutche For\-sch\-ungs\-gemeinschaft,
and the Russian Foundation for Basic Research}}

\author{\hnl{E.A.~Kolganova}{{http://thsun1.jinr.ru/~kea}}\footnote{On leave of absence from
Bogoliubov Laboratory of Theoretical Physics,  Joint
Institute for Nuclear Research, Joliot-Curie 6, 141980~Dubna, Russia},
\hnl{Y.K.~Ho}{http://www.iams.sinica.edu.tw/introa/research_team/cl504.htm}}

\address{Institute of Atomic and Molecular Sciences,
Academia Sinica\\ P.O.Box 23-166, Taipei, Taiwan 106, ROC}

\author{\hnl{A.K.~Motovilov}{http://thsun1.jinr.ru/~motovilv}}
\address{Bogoliubov
Laboratory of Theoretical Physics\\ Joint Institute for
Nuclear Research
\\ Joliot-Curie 6, 141980~Dubna,  Russia}

\author{\hnl{W. Sandhas}{http://www.th.physik.uni-bonn.de/th/People/sandhas/}}
\address{Physikalisches Institut der Universit\"at Bonn\\
Endenicher Allee 11-13, D-53115 Bonn, Germany}

\bigskip

\date{April 11, 2003}


\begin{abstract}
\phantom{i}

\bigskip

\medskip

We apply a hard-core version of the Faddeev differential
equations to the  $^3$He$^4$He$_2$ three-atomic system.
Employing the TTY interatomic potential by Tang, Toennies
and Yiu we calculate the binding energy of the
$^3$He$^4$He$_2$ trimer and the scattering length of a
$^3$He atom off a $^4$He dimer.

\pacs{02.60.Nm, 21.45.+v, 34.40.-m, 36.40.+d}
\end{abstract}

\maketitle

 \section{Introduction}
 \label{SIntro}

 Small Helium clusters attract a lot of attention in
 various fields of physical chemistry and molecular
 physics. There is a great number of experimental and
 theoretical studies of the $^4$He three-atomic system
 (see, e.\,g., \cite{GrebToeVil}-- \cite{Roudnev1} and
 references cited therein). The non-symmetric system
 $^3$He$^4$He$_2$ found comparatively little attention. To
 the best of our knowledge the $^3$He$^4$He$_2$ trimers
 were studied numerically only in Refs.
 \cite{EsryLinGreene}, \cite{Nielsen},  \cite{Bressani},
 \cite{Roudnev}. Except Ref. \cite{Roudnev}, there are
 still no scattering calculations reported for this system.

The present work is a sequel to the investigation of the
helium three-atomic systems undertaken in Refs.
\cite{KMS-JPB}, \cite{MSSK},  \cite{CzJP} based on a
mathematically rigorous hard-core version of the Faddeev
differential equations. This method allows one to overcome
in particular the strong-repulsion problem arising in
examination of atomic systems. Along the same line we now
investigate the $^3$He$^4$He$_2$ bound states as well as
the scattering of a $^{3}$He atom off a $^4$He dimer at
ultra-low energies.

For the moment we restrict ourselves to the use of the TTY
interatomic He--He potential by Tang, Toennies and Yiu
\cite{Tang95}. Computations with other realistic He--He
interactions are in progress and will be reported
elsewhere.

 \section{Method}

 In this section we only describe the main features of the
 method employed. Further technical details can be found in
 \cite{KMS-JPB},  \cite{MSSK},  \cite{CzJP}.

 It is assumed that the $^3$He$^4$He$_2$ three-atomic
 system is in the state with the total angular momentum
 $L = 0$. Then the partial angular analysis reduces the
 initial hard-core Faddeev equations (see, e.g.,
 \cite{KMS-JPB}) to a system of coupled two-dimensional
 integro-differential equations
 \begin{eqnarray}
 \nonumber
 &{\left(-\displaystyle\frac{\partial^2}{{\partial x}^2}-
 \frac{\partial^2}{{\partial y}^2}+
 l(l+1)\left(\displaystyle\frac{1}{x^2}+\frac{1}{y^2}\right)-E
 \right)f^{(\alpha)}_l(x,y)}\\
 \label{Fparts}
 & \qquad=\left\{\begin{array}{cc} 0, & x<c \\
 -V_\alpha(x)\psi_l^{(\alpha)}(x,y), & x>c
 \end{array}\right., \qquad \alpha=1,2,3,
 \end{eqnarray}
 Here, $x,y$ stand for the reduced Jacobi variables, and
 $c$ for the core radius which for simplicity is taken the
 same for all three interatomic interactions. The He--He
 interaction potentials $V_{\alpha}(x)$ are assumed to be
 central. The partial wave functions $\psi_l^{\alpha}(x,y)$
 are related to the partial wave Faddeev components
 $f^{(\alpha)}_l(x,y)$ by
 \begin{equation}
 \label{psi}
 \psi_l^{(\alpha)}(x,y)=
 f_l^{(\alpha)}(x,y)+\displaystyle\sum\limits_{l'}
 \sum\limits_{\beta\neq\alpha}\int\limits_0^1 d\eta \,
 h^0_{(\alpha; ll0)(\beta; l'l'0)}(x,y,\eta)
 f^{(\beta)}_{l'}(x_{\beta\rightarrow\alpha}(\eta),
 y_{\beta\rightarrow\alpha}(\eta)),
 \end{equation}
 where $1 \leq{\eta}\leq 1$. The explicit form of the
 function $h_{(\alpha;ll0)(\beta;l'l'0)}$ can be found in
 Refs.~\cite{CzJP}, \cite{MF}. The functions $f^{(\alpha)}_l(x,y)$
 satisfy the hard-core boundary conditions
 \begin{equation}
 \label{HCparts}
 f_{l}^{(\alpha)}(x,y)\left.\right|_{x=0}
      =f_{l}^{(\alpha)}(x,y)\left.\right|_{y=0}=0\,
      \quad {\rm and} \quad
 \left.\psi_l^{(\alpha)}(x,y)\right|_{x=c}=0,\qquad \alpha=1,2,3.
 \end{equation}
 Further, the system (\ref{Fparts})--(\ref{HCparts}) is
 supplemented with the corresponding asymptotic boundary
 conditions for $f_{l}^{(\alpha)}(x,y)$ as $x\to\infty$
 and/or $y\to\infty$ (see \cite{KMS-JPB}, \cite{CzJP},
 \cite{MF}).

 Here we only deal with a finite number of equations
 (\ref{Fparts})--(\ref{HCparts}), assuming that $l\leq
 l_{\rm max}$ where $l_{\rm max}$ is a certain fixed
 number.

 \section{Results}

 First, we employed the equations (\ref{Fparts}) --
 (\ref{HCparts}) and the corresponding bound-state
 asymptotic boundary conditions to calculate the binding
 energy of the helium trimer $^3$He$^4$He$_2$. Recall that
 in this work as a He-He interaction we used the TTY
 potential by Tang, Toennies and Yiu \cite{Tang95}.
 As in \cite{KMS-JPB}, \cite{MSSK}, \cite{CzJP} we used a
 finite-difference approximation of the boundary-value
 problem (\ref{Fparts})--(\ref{HCparts}) in the polar
 coordinates $\rho = \sqrt{x^2+y^2}$ and
 $\theta = \arctan(y/x)$. The grids were chosen such that the
 points of intersection of the arcs $\rho = \rho_i$,
 $i = 1,2,\ldots, N_\rho$ and the rays $\theta = \theta_j$,
 $j = 1,2,\ldots, N_\theta$ with the core boundary $x = c$
 constitute the knots. The value of the core radius  was
 chosen to be $c = 1$\,{\AA} by the same argument as in
 \cite{MSSK}. Also the idea for choosing the grid radii
 $\rho_i$ (and, thus, the grid hyperangles $\theta_j$) was
 the same as described in \cite{MSSK}.

\begin{table}[h]
\caption{\label{tableTrimerGS}Absolute value of
the $^3$He$^4$He$_2$ trimer binding energies (in mK) for
the TTY potential.}
\begin{center}
\begin{tabular}{cccc}
\hline\hline\hline
& & &  \\[-2ex]
$\qquad l_{\rm max}\qquad$ & $\quad$This work$\quad$ &
$\quad$Ref. \cite{Bressani}$\quad$ &
$\quad$Ref. \cite{Roudnev}$\quad$ \\
& & &  \\[-2ex]
\hline\hline & & &  \\[-2ex] 0 &  $7.25$ &   & \\
        \cline{1-2}
        & & &  \\[-2ex]
2 & $13.09$ & & \\
        \cline{1-2}
        & & &  \\[-2ex]
4 & $13.78$ & $14.165$ & $14.1$ \\
\hline\hline\hline
\end{tabular}
\end{center}
\end{table}

We assumed that $\hbar^2/{\rm m} = 12.12$\,K\AA$^2$ where
${\rm m}$ stands for the mass of a $^4{\rm He}$ atom. The
mass ratio ${\rm m_{^3{\rm He}}}/{\rm m_{^4{\rm He}}}$ was
assumed to be equal to $0.753517$. Notice that for the TTY
potential the $^4$He-dimer energy is 1.30962\,mK
\cite{MSSK}.

\begin{table}[h]
 \caption{\label{tableTrimerLen}Estimations for the
 $^{3}$He atom\,--\,$^4$He dimer scattering length (in
 \AA) with the TTY potential.}
 \begin{center}
 \begin{tabular}{ccc}
 \hline\hline\hline
 & &  \\[-2ex]
 $\qquad l_{\rm max}\qquad$ & $\quad$This work$\quad$
 & $\quad$Ref. \cite{Roudnev}$\quad$ \\
 \hline\hline
 & &  \\[-2ex]
 0  & $38.8$ &   \\
 2  & $22.4$ &     \\
 4  & $21.2$  &  19.6  \\
 \hline\hline\hline
 \end{tabular}
 \end{center}
\end{table}

The best dimensions of the grids which we employed in this
investigation were $N_\rho = 600$ and $N_\theta = 605$
with cut-off hyperradius $\rho_{\rm max} = 200$\,{\AA}.
Our results for the $^3$He$^4$He$_2$ binding energy
obtained for a grid with such $N_\rho$, $N_\theta$, and
$\rho_{\rm max}$, as well as the results available in the
literature, are presented in Table \ref{tableTrimerGS}. Is
is found that most of the contribution to the binding
energy stems from the $l=0$ and $1\leq l\leq 2$ partial
components, about 53\% and 42\%, respectively. The overall
contribution from the $l=3$ and $l=4$ partial wave
components is of the order of 5\,\%. A certain (but rather small)
deepening of the $^3$He$^4$He$_2$ binding energy  may also
be expected due to choosing the grids with larger
$N_\theta$ and $N_\rho$.

Being a more light particle than $^4$He, the $^3$He atom
does not form a bound molecule with the $^4$He counterpart
and no $^3$He dimer exists. As a consequence,
$^3$He$^4$He$_2$ is a more loosely bound system than the
$^4$He trimer (see, e.g. \cite{KMS-JPB} and \cite{MSSK}).
In partiuclar, there is no exited state of the
$^3$He$^4$He$_2$ system in contrast to the symmetric
$^4$He$_3$ trimer that is well known to have an exited
state of Efimov nature.

 Our results for the $^3$He--$^4$He$_2$ scattering length
 calculated also for a grid with $N_\rho=600$,
 $N_\theta=605$, and $\rho_{\rm max}=200$\,{\AA} are shown
 in Table \ref{tableTrimerLen}.

 \section*{Acknowledgements}
 \bigskip
 {The authors are grateful to Prof.~V.\,B.\,Belyaev and
 Prof.~H.\,Toki for making us possible to perform calculations at
 the supercomputer of the Research Center for Nuclear Physics of
 Osaka University, Japan.}

\end{document}